\documentclass{PoS}

\usepackage{amsmath}    
\usepackage{graphicx}   
\usepackage{verbatim}   
\usepackage{xspace}

\usepackage{amssymb}

\usepackage{ifthen}

\usepackage{bm}
\usepackage{esvect}
\usepackage{gensymb} 
\usepackage[ddmmyyyy]{datetime} \longdate

\usepackage[style=numeric-comp,maxnames=3,backend=bibtex8,doi=false,isbn=false,url=false,sorting=none,sortcites=true]{biblatex}
\AtEveryBibitem{\clearfield{title}}

\usepackage[caption=false]{subfig}
\DeclareCaptionListOfFormat{myparens}{#2}
\newcommand{\Subref}[1]{\protect\subref{#1}}

\definecolor{figColBox}{rgb}{1,1,1}

\usepackage{xcolor}
\definecolor{white}{rgb}{1,1,1}
\definecolor{red}{rgb}{1,0,0}
\definecolor{blue}{rgb}{0,0,1}
\definecolor{darkBlue}{rgb}{0,0,.4}
\definecolor{green}{rgb}{0,1,0}
\definecolor{magenta}{rgb}{1,0,.6}
\definecolor{lightblue}{rgb}{0,.5,1}
\definecolor{lightpurple}{rgb}{.6,.4,1}
\definecolor{gold}{rgb}{.6,.5,0}
\definecolor{orange}{rgb}{1,0.4,0}
\definecolor{hotpink}{rgb}{1,0,0.5}
\definecolor{newcolor2}{rgb}{.5,.3,.5}
\definecolor{newcolor}{rgb}{0,.3,1}
\definecolor{newcolor3}{rgb}{1,0,.35}
\definecolor{darkgreen1}{rgb}{0, .35, 0}
\definecolor{darkgreen}{rgb}{0, .6, 0}
\definecolor{darkred}{rgb}{.75,0,0}
\xdefinecolor{olive}{cmyk}{0.64,0,0.95,0.4}
\xdefinecolor{purpleish}{cmyk}{0.75,0.75,0,0}
\definecolor{grayBold}{rgb}{0,0,0}
\definecolor{battleshipgrey}{rgb}{0.52, 0.52, 0.51}
\definecolor{cinereous}{rgb}{0.6, 0.51, 0.48}
\definecolor{darkgray}{rgb}{0.66, 0.66, 0.66}
\definecolor{davysgrey}{rgb}{0.33, 0.33, 0.33}
\definecolor{dimgray}{rgb}{0.41, 0.41, 0.41}
\definecolor{gainsboro}{rgb}{0.86, 0.86, 0.86}
\definecolor{grullo}{rgb}{0.66, 0.6, 0.53}
\definecolor{manatee}{rgb}{0.59, 0.6, 0.67}
\definecolor{oldlavender}{rgb}{0.47, 0.41, 0.47}
\definecolor{oldmauve}{rgb}{0.4, 0.19, 0.28}
\definecolor{khaki}{rgb}{0.76, 0.69, 0.57}
\definecolor{paynesgrey}{rgb}{0.25, 0.25, 0.28}
\definecolor{platinum}{rgb}{0.9, 0.89, 0.89}
\definecolor{whitesmoke}{rgb}{0.96, 0.96, 0.96}
\definecolor{coolblack}{rgb}{0.0, 0.18, 0.39}
\definecolor{royalblue}{rgb}{0.25, 0.41, 0.88}

\definecolor{red0}{RGB}{243,61,99}
\definecolor{red1}{RGB}{198,40,40}
\definecolor{green0}{RGB}{139,195,74}
\definecolor{blue0}{RGB}{33,150,243}
\definecolor{blue1}{RGB}{57,73,171}
\definecolor{blue2}{RGB}{134, 165, 216}
\definecolor{blue3}{RGB}{14,117,194}
\definecolor{blue4}{RGB}{10,86,143}

\definecolor{lightblack}{RGB}{55,59,61}

\definecolor{blueTbl}{RGB}{17, 143, 237}
\definecolor{greenTbl}{RGB}{137, 197, 65}

\definecolor{blueTblFaint}{RGB}{207, 232, 252}
\definecolor{greenTblFaint}{RGB}{231, 243, 216}

\definecolor{grayBckCol}{RGB}{244, 244, 244}
\definecolor{faintGrayBckCol}{RGB}{246, 246, 246}

\definecolor{figColBox}{rgb}{1,1,1}
\definecolor{linkCol}{RGB}{59,80,125}

\newcommand{\eg}{e.g.\/,\xspace}
\newcommand{\Eg}{E.g.\/,\xspace}

\newcommand{\insitu}{{in-situ}\xspace}

\newcommand{\Sim}{\sim\kern-0.2em\xspace}

\newcommand{\mrm}[1]{\ensuremath{\mathrm{#1}}}
\newcommand{\txt}[1]{\text{#1}}
\newcommand{\tit}[1]{\textit{#1}}

\newcommand{\powB}[1]{\ensuremath{{10^{{#1}}}}\xspace}  

\newcommand{\gamlib}{\textit{GammaLib}\xspace}
\newcommand{\ctools}{\textit{ctools}\xspace}
\newcommand{\prodIrf}{\textit{prod3b-v1}\xspace}
\newcommand{\tensorflow}{\textit{tensorflow}\xspace}

\newcommand{\gev}{\ensuremath{\txt{GeV}}\xspace}

\newcommand{\erg}{\ensuremath{\txt{erg}}\xspace}

\newcommand{\dgr}{\ensuremath{\degree}\xspace}

\newcommand{\scnd}{\ensuremath{\txt{s}}\xspace}

\newcommand{\hess}{{H.E.S.S.}\xspace}
\newcommand{\magic}{{MAGIC}\xspace}
\newcommand{\veritas}{{VERITAS}\xspace}
\newcommand{\fermi}{{\textit{Fermi}}\xspace}
\newcommand{\fermil}{{\textit{Fermi}-LAT}\xspace}

\newcommand{\cta}{{CTA}\xspace}

\newcommand{\roi}{RoI\xspace}

\newcommand{\fov}{FoV\xspace}

\newcommand{\grb}{\ensuremath{\txt{GRB}}\xspace}
\newcommand{\grbs}{\ensuremath{\txt{GRBs}}\xspace}

\newcommand{\llgrbs}{\ensuremath{\txt{LL-GRBs}}\xspace}

\newcommand{\hlgrbs}{\ensuremath{\txt{high-luminosity GRBs}}\xspace}

\newcommand{\ebl}{\ensuremath{\txt{EBL}}\xspace}
\newcommand{\liso}{\ensuremath{{L_{\gamma,\text{iso}}}}\xspace}

\newcommand{\gamray}{\ensuremath{\gamma\txt{-ray}}\xspace}

\newcommand{\irfs}{IRFs\xspace}

\newcommand{\plaw}{\ensuremath{\txt{PL}}\xspace}

\newcommand{\lstm}{LSTM\xspace}
\newcommand{\lstms}{LSTMs\xspace}
\newcommand{\rnn}{RNN\xspace}
\newcommand{\rnns}{RNNs\xspace}

\newcommand{\andet}{anomaly detection\xspace}
\newcommand{\Andet}{Anomaly detection\xspace}
\newcommand{\cls}{classification\xspace}

\newcommand{\ts}{\ensuremath{\mrm{TS}}\xspace}

\newcommand{\tsCls}{\ensuremath{\mrm{TS}_{\mrm{clas}}}\xspace}

\newcommand{\predCls}{\ensuremath{\zeta}\xspace}

\newcommand{\predClsB}{\ensuremath{\zeta_{\mrm{bck}}}\xspace}
\newcommand{\predClsS}{\ensuremath{\zeta_{\mrm{sig}}}\xspace}

\newcommand{\pdet}{\ensuremath{p_{\mrm{det}}}\xspace}
\newcommand{\fdet}{\ensuremath{f_{\mrm{det}}}\xspace}

\title{Deep learning detection of transients}

\ShortTitle{Deep learning detection of transients}

\author{\speaker{Iftach Sadeh}\\
        DESY, Platanenallee 6, 15738 Zeuthen, Germany\\
        E-mail: \email{iftach.sadeh@desy.de}}

\abstract{
The next generation of observatories will facilitate the discovery of new types of
astrophysical transients. The detection of such phenomena, whose characteristics are
presently poorly constrained, will hinge on the ability to perform blind searches. We present
a new algorithm for this purpose, based on deep learning. We incorporate two approaches,
utilising anomaly detection and classification techniques. The first is model-independent,
avoiding the use of background modelling and instrument simulations. The second method
enables targeted searches, relying on generic spectral and temporal patterns as input. We
compare our methodology with the existing approach to serendipitous detection of gamma-ray transients.
We use our framework to derive the detection prospects of low-luminosity
gamma-ray bursts with the upcoming Cherenkov Telescope Array.
Our method is an
unbiased, data-driven approach for multiwavelength and multi-messenger
transient detection.
}

\FullConference{36th International Cosmic Ray Conference -ICRC2019-\\
        July 24th - August 1st, 2019\\
        Madison, WI, U.S.A.}

\begin{document}

\section{Introduction}
%
  Transient astrophysical events at
  high energies enable the study
  of a broad range of fundamental phenomena.
  We present a new method,
  intended for the detection of such events.
  Our algorithm is based on deep learning,
  using recurrent neural networks. It employs
  two complementary approaches, optimised for both
  model-independent and targeted searches,
  respectively denoted as
  \tit{anomaly detection}, and
  \tit{classification}.
  Our approach is by design generic.
  The methodology is not restricted to a specific energy regime,
  type of input, or time scale.
  In particular, it is well suited for multiwavelength
  and multi-messenger searches,
  where different observables are combined.
  It can therefore easily be adapted
  for many types of transient searches.
  We illustrate the method
  for the case of serendipitous detection of
  low-luminosity gamma ray (\gamray) bursts
  (\grbs; \llgrbs) with \cta, the upcoming
  Cherenkov Telescope Array.\footnote{~\cta:~\url{https://www.cta-observatory.org/}.}

\section{New detection methods}
%
  %
  Machine learning is widely used in
  astronomy~\cite{Sadeh:2015lsa, 2017APh....89....1K,2018MNRAS.476.2117R, 2018APS..APRL01031K, 2019MNRAS.484...93D, 2018MNRAS.476.5365S}.  
  In the current study, we utilise
  a recurrent neural network (\rnn),
  made up of long short-term memory (\lstm) units.      
  \rnns are a type of artificial neural network,
  which is well suited for time series
  analysis~\cite{Sutskever:2014:SSL:2969033.2969173, Shen:2017jkj}.
  For a review of deep learning and \rnns, see~\cite{deepLearningReview}.

\begin{figure*}[tp]
    \begin{minipage}[c]{1\textwidth}

      \begin{minipage}[c]{1\textwidth}
        \begin{center}
          \colorbox{figColBox}{\includegraphics[trim=16mm 90mm 49mm 40mm,clip,width=1\textwidth]{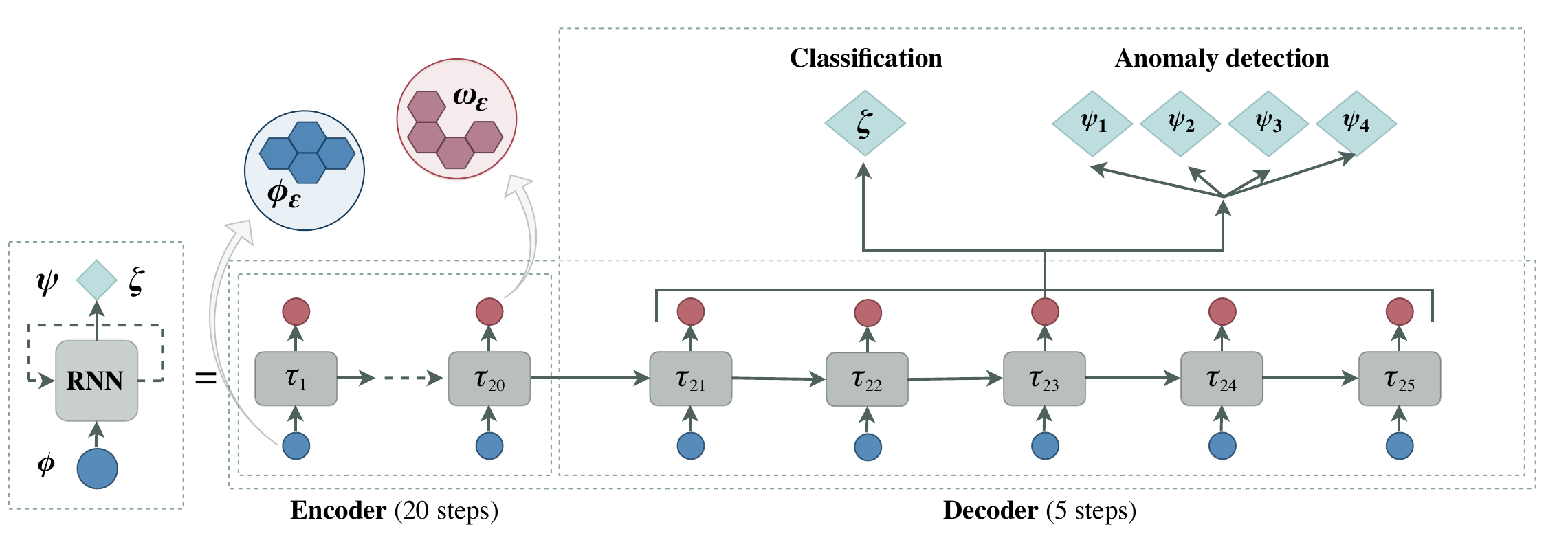}}
        \end{center}
      \end{minipage}\hfill
      %
      %
      \begin{minipage}[c]{1\textwidth}
        \begin{center}
          \begin{minipage}[t]{1\textwidth}\begin{center}
            \caption{\label{FIGannArch}Schematic
              design of the \rnn used in this study.
              The network may be decomposed into
              an encoder and a decoder, respectively
              representing~$20$ and~$5$ time steps, $\tau$,
              of \lstm units (rectangles).
              The input data, $\phi$, (blue circles)
              make up $4$ numbers for each time
              step (blue hexagons), corresponding to \gamray counts in
              different energy bins, $\epsilon$.
              The direct output of the \lstms,
              $\omega_{\epsilon,\tau}$, (red hexagons) are predictions
              for event counts for each step and energy bin.
              The output of the \rnn (diamonds)
              is different for each implementation.
              For \andet, $\psi_{1-4}$, provide the predicted
              background counts per energy bin,
              integrated over the decoder time steps.
              The output of the \cls method, $\zeta$,
              is used to derive a probability density function.}
          \end{center}\end{minipage}\hfill
        \end{center}
      \end{minipage}\hfill
      \vspace{15pt}
    \end{minipage}\hfill
  \end{figure*}

  We utilize the open-source
  software, \tensorflow,
  for our implementation~\cite{tensorflow2015-whitepaper}.
  The architecture of the \rnn is
  illustrated in Figure~\ref{FIGannArch}.
  The network accepts an input which
  corresponds to $25$~time steps, each representing
  a~${1~\scnd}$ interval of \gamray data.
  The different steps are implemented as \rnn cells. A cell
  is composed of a pair of \lstm layers,
  respectively comprising~$128$ and~$64$
  hidden units in the current implementation. 
  The network may be decomposed into an \tit{encoder} and a
  \tit{decoder}. The encoder receives $20$~time steps
  as input.
  A potential transient signal event is then
  searched for within the final $5$~time steps.

  %
  %

  \Andet represents a model-independent
  approach, where transient events are identified based on
  their divergence from the expected background.
  This simple methodology is completely data-driven, and
  is able to adapt to real-time evolution of the background.
  \Andet is decoupled from instrument simulations.
  It is insensitive to uncertainties on
  modelling of the background,
  the atmosphere, or other observational artefacts which do not
  evolve strongly in time.
  In the current example,
  Poissonian statistics are assumed
  for both the background and the signal
  models.\footnote{~For brevity, we refer to the
  background-only hypothesis as the \tit{background model},
  and to the background$+$signal hypothesis
  as the \tit{signal model}.}
  The background in this case is derived \insitu,
  using data exclusively from within the
  \roi for the source.
  The network is trained using background-only 
  events for all $25$~time steps.
  The test statistic for detecting a transient is based on
  the $5$~decoder time steps. It encapsulates
  the difference between the counts predicated by the \rnn, and
  actual observations.

  Our second method employs classification.
  In this case, we
  train an estimator to identify transient patterns.
  The \cls approach increases the sensitivity of specific searches,
  such as for \llgrbs. For instance,
  the time structure of transient events is
  naturally incorporated as a part
  of the training process,
  avoiding the need for explicit modelling.
  Further more, simple training examples are sufficient for
  subsequent detection of more complicated intrinsic
  spectra, as discussed below.
  The network is trained using labelled examples
  of background and signal events.

  The output of the \rnn, $\zeta$, is the inferred
  classification metric for a given event,
  (see Figure~\ref{FIGclsDist} below).
  The test statistic for identifying
  a signal event is derived from
  the distributions of classified 
  signal and background events, respectively
  denoted as \predClsS and \predClsB.
  It is defined as,
  \begin{equation}
  \ts = -2 \log \left( \frac{\predClsB} {\predClsS} \right) \, ,
  \label{eqGrbPlModel} \end{equation}
  following the prescription of Ref\/.~\cite{Cranmer:2015bka}.

  For additional details on these methods, see Ref\/.~\cite{Sadeh:2019qbb}.

\section{Transient simulations}
  %
    %
    One of the interesting source populations that might
    be fully unveiled in the near future is that of
    low-luminosity \gamray bursts~\cite{Liang:2006ci, Virgili:2008gp}.
    \llgrbs are distinguished by
    low isotropic equivalent luminosities,
    generally,
    ${\powB{46} < \liso < \powB{49}~\erg\,\scnd^{-1}}$.
    A sub-class of the population
    of long \gamray bursts (\grbs), they have been connected
    to mildly relativistic supernovae~\cite{Cano:2016ccp},
    and are potential sources of of ultra high-energy
    cosmic rays and
    neutrinos~\cite{2008PhRvD..78b3005M, 2018PhRvD..97h3010Z, 2018arXiv180807481B}.
    There are indications that the observable rate of \llgrbs in the
    local Universe (redshift, ${z < 0.1}$)
    is high, of the order of
    ${200\,\mrm{Gpc}^{-3}\,\mrm{yr}^{-1}}$~\cite{Sun:2015bda}.
    They are thus appealing targets for blind real-time searches.
    Observation of \llgrbs is challenging
    with the current generation of
    ground- and space-based
    observatories\footnote{~\Eg~\fermi:~\url{https://fermi.gsfc.nasa.gov/}; \hess:~\url{https://www.mpi-hd.mpg.de/hfm/HESS/HESS.shtml}; \magic:~\url{https://wwwmagic.mpp.mpg.de/}; \veritas:~\url{https://veritas.sao.arizona.edu/}.}~\cite{Sun:2015bda, Abdollahi:2016rso}.
    However, the upcoming \cta
    will significantly enhance their detection prospects,
    allowing \gamray measurements down to~${\Sim20~\gev}$,
    within a large field-of-view (\fov).

    We use the following as reference events for simulating
    the possible \gamray signatures of \llgrbs:
    GRBs~${\mrm{080916C}}$, ${\mrm{090323}}$,
     ${\mrm{090510}}$, ${\mrm{090902B}}$,
     and~${\mrm{110731A}}$. 
    These are bright \hlgrbs, which have been detected at
    high energies with \fermil (see Ref\/.~\cite{Ackermann:2013zfa}
    and references therein).
    We assume that the \gev emission of these bursts
    is a simple power law (\plaw)
    extension of a Band-like model to high energies~\cite{1993ApJ...413..281B}.
    We only consider
    those bursts which exhibit durations of the order of
    tens of seconds. We thus exclude the
    population of ultra-long \grbs, which might
    involve unique emission mechanisms,
    such as shock breakouts~\cite{Waxman:2007rr, Levan:2013gcz}.

    We randomly
    shift the reference \grbs in redshift and luminosity
    to the expected ranges for \llgrbs,
    scaling the observed flux accordingly~\cite{Inoue:2013vy}.
    In order
    to simulate the signals at \gev energies,
    we nominally assume a simple spectral/temporal \plaw model,
    \begin{equation}
    M_{\mrm{PL}}(E,t) =
      k_{0}
      \left(\frac{E}{E_{0}}\right)^{-\Gamma}
      t^{-\tau} .
    \label{eqGrbPlModel} \end{equation}
    \noindent
    The \tit{prefactor} and \tit{pivot energy},
    $k_{0}$ and $E_{0}$, are derived directly from the
    flux of the \grb.
    The spectral index, $\Gamma$, and temporal
    decay index, $\tau$, are randomly selected
    for each event, uniformly distributed
    within
    $1.9 < \Gamma < 2.7$
    and
    $0.8 < \tau < 2$.
    These properties
    generally correspond to the expectations
    for the low-luminosity population.

    We generate \cta
    events using the open-source software,
    \ctools,~\cite{Knodlseder:2016nnv}
    simulating the Northern array~\cite{Acharya:2017ttl},
    using the publicly available \irfs (version \prodIrf).
    We exclusively use \irfs
    optimised for ${30~\min}$ observations at zenith
    angles of~${\Sim20\dgr}$.
    The region of interest (\roi)
    for the simulation is chosen as a circular
    region with a radius of ${0.25\dgr}$. The circle is centred 
    at the position of the source, which
    is displaced by ${0.5\dgr}$ from the centre
    of the \fov.

    The inputs to the \rnn corresponding to a given time step 
    are event counts in $4$
    logarithmically-spaced energy bins within
    ${30 < E_{\mrm{\gamma}} < 200~\gev}$.
    The inputs to the encoder
    are assumed to correspond to background-only
    counts in all cases.
    The input to the decoder and the
    output of the network depend on the
    type of inference being used.
    We conduct searches over the ${5~\scnd}$ intervals
    that coincide with the beginning of bursts in the signal sample.
    We assume a conservative correction for trials~\cite{BILLER1996285},
    accounting for $100$~h of observations
    at ${1~\scnd}$ search intervals.

    We explore the effects of the 
    extragalactic background light (\ebl) on the observed spectra.
    Comparing different models~\cite{Franceschini:2008tp, doi:10.1111/j.1365-2966.2010.17631.x, doi:10.1111/j.1365-2966.2012.20841.x},
    we find that the \ebl has little effect on the \llgrbs
    in our simulation, due primarily to their low redshift.

\section{Results}
\begin{figure*}[tp]

      \begin{minipage}[c]{1\textwidth}

        \begin{minipage}[c]{1\textwidth}
          \begin{center}
            \begin{minipage}[c]{0.47\textwidth}\begin{center}
              \colorbox{figColBox}{\includegraphics[trim=0mm 12mm 0mm 18mm,clip,width=.98\textwidth]{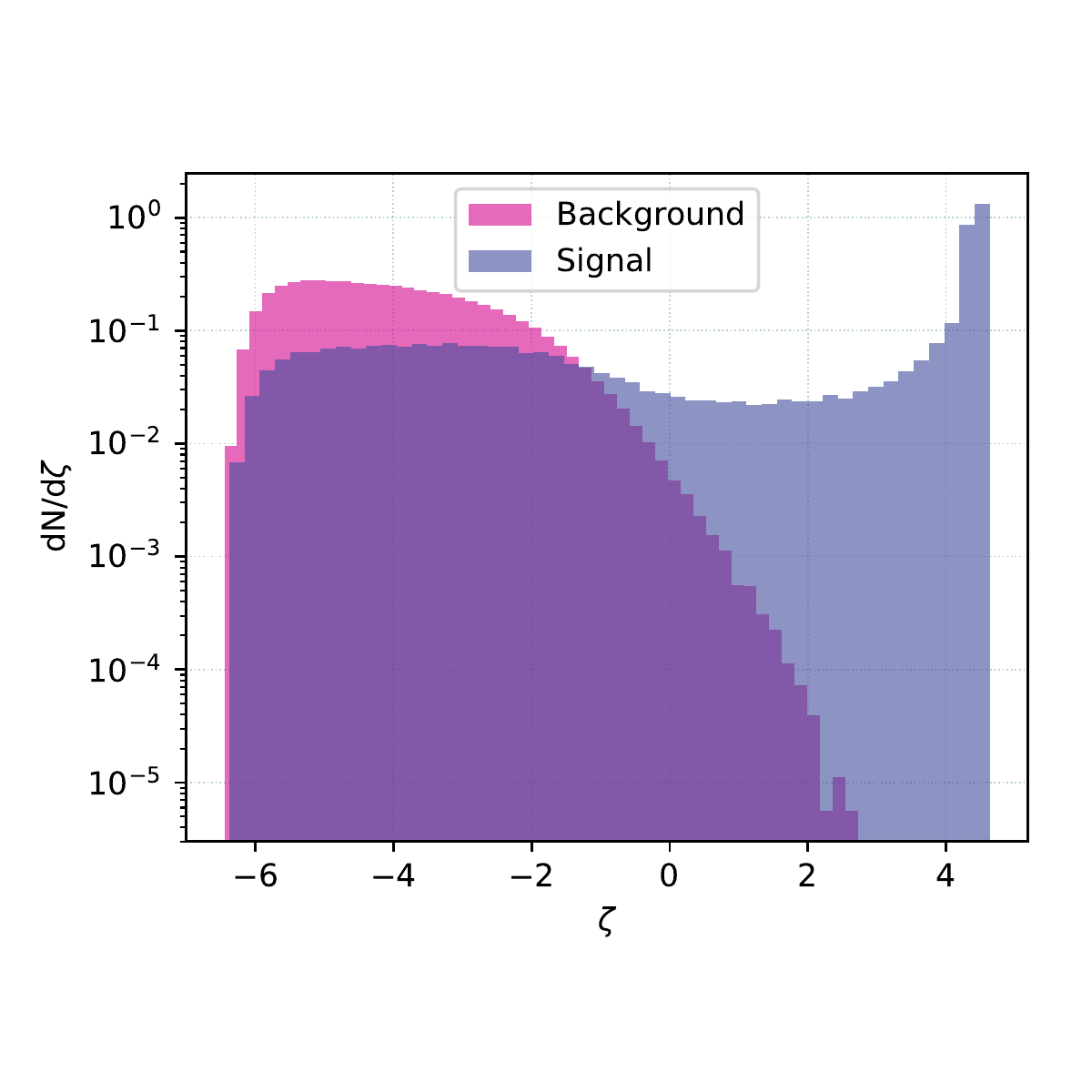}}
            \end{center}\end{minipage}\hfill
            \begin{minipage}[c]{0.47\textwidth}\begin{center}
              \colorbox{figColBox}{\includegraphics[trim=0mm 12mm 0mm 18mm,clip,width=.98\textwidth]{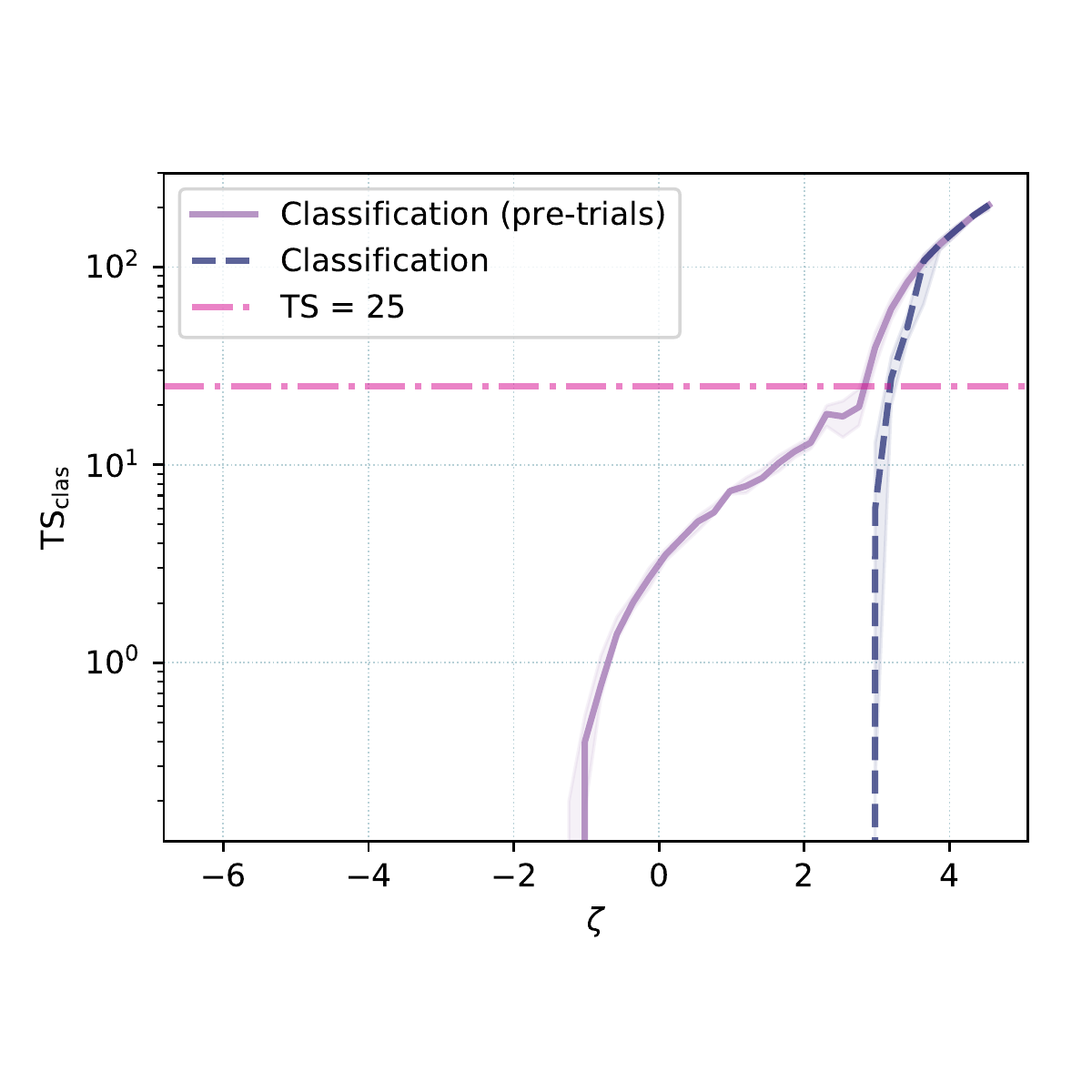}}
            \end{center}\end{minipage}\hfill 
          \end{center}
        \end{minipage}\hfill
        \vspace{5pt}
        \begin{minipage}[c]{1\textwidth}
          \begin{center}
            \begin{minipage}[c]{0.47\textwidth}\begin{center}
              \centering\subfloat[]{\label{FIGclsDist}}
            \end{center}\end{minipage}\hfill
            \begin{minipage}[c]{0.47\textwidth}\begin{center}
              \centering\subfloat[]{\label{FIGclsTs}}
            \end{center}\end{minipage}\hfill 
          \end{center}
        \end{minipage}\hfill
        %
        %
        \begin{minipage}[c]{1\textwidth}
          \begin{center}
            \begin{minipage}[t]{1\textwidth}\begin{center}
              \caption{\label{FIGclsPerformance}Parametrisation
                of the performance of the trained \cls method.
                \Subref{FIGclsDist}~Distributions of the
                \cls metric, \predCls, for the signal
                and background samples, as indicated.
                \Subref{FIGclsTs}~The parametrised
                \cls test statistic, \tsCls,
                as a function of \predCls,
                before and after the correction for trials.
                The dashed-dotted horizontal line highlights the
                value, $\ts = 25$.}
            \end{center}\end{minipage}\hfill
          \end{center}
        \end{minipage}\hfill
        \vspace{15pt}
      \end{minipage}\hfill

    \end{figure*}

  %
    Using the trained \rnn in the \cls mode,
    we derive the distributions of the
    \cls metric, \predCls, for the
    background and signal samples,
    as shown in Figure~\ref{FIGclsDist}.
    The corresponding relation between
    the \cls test statistic, \tsCls, and \predCls
    is presented in Figure~\ref{FIGclsTs}.
    We evaluate our results using
    the \tit{detectability} metric,
    ${\pdet = \left< \rho_{\mrm{det}} \right>}$, defined 
    event-by-event for
    \begin{equation}
      \begin{matrix}
        %
        \rho_{\mrm{det}}(t) = 
        &
        \left\{ \begin{matrix} 
        0
        \;,
        &
        t < \ts_{5\sigma}
        \\
        1\;,
        &
        t \geq \ts_{5\sigma}
        \end{matrix} \right. .
        & \\
      \end{matrix}
    \label{eqGrbDetectability} \end{equation}
    \noindent
    Here $t$ represents the test
    statistic derived for a given detection method;
    ${\ts_{5\sigma}}$ is the corresponding
    threshold for a $5\sigma$ detection,
    where \eg for a model with a single degree
    of freedom, ${\ts_{5\sigma} = 25}$~\cite{Wilks:1938dza}.

\begin{figure*}[tp]

      \begin{minipage}[c]{1\textwidth}

        \begin{minipage}[c]{1\textwidth}
          \begin{center}
            \begin{minipage}[c]{0.47\textwidth}\begin{center}
              \colorbox{figColBox}{\includegraphics[trim=0mm 12mm 0mm 18mm,clip,width=.98\textwidth]{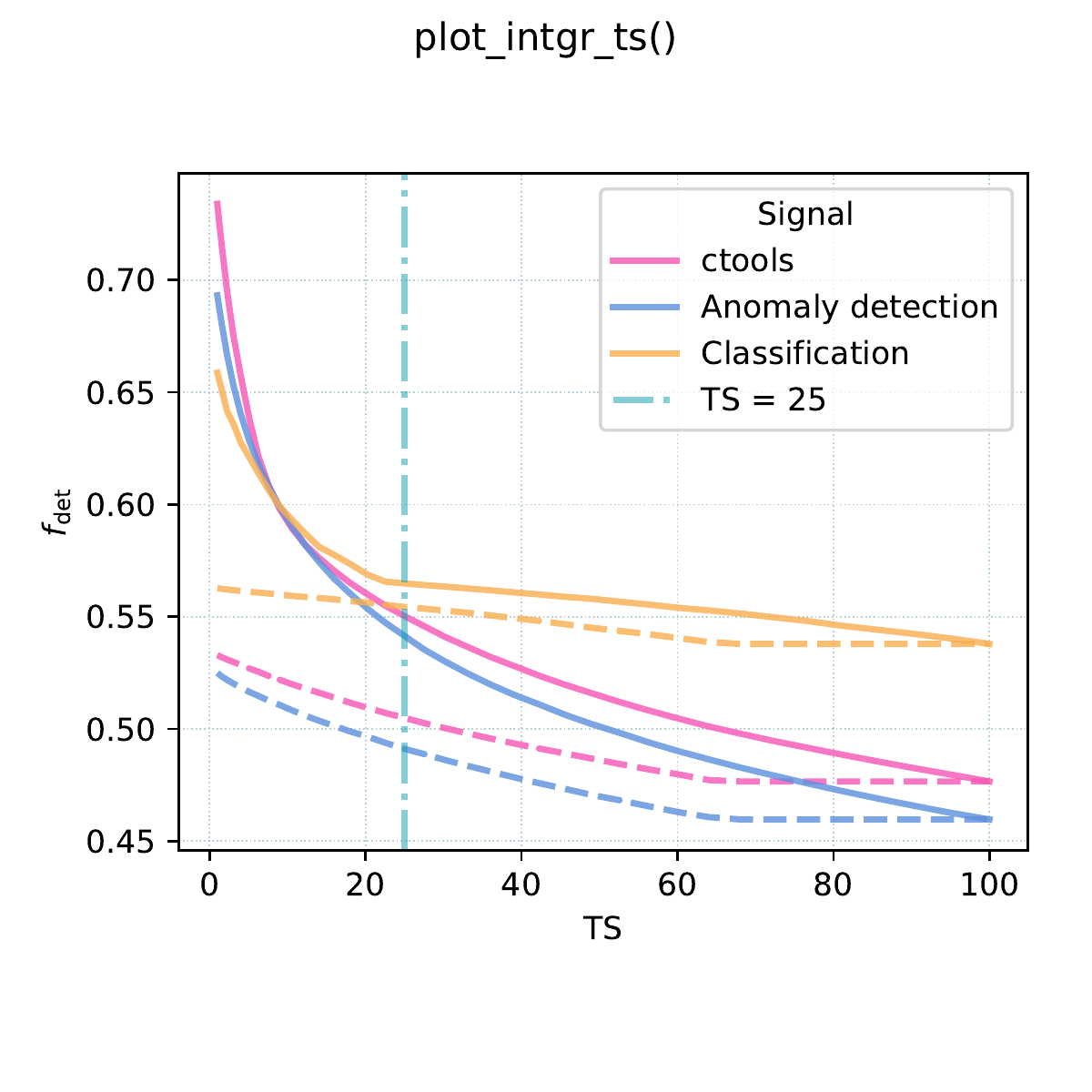}}
            \end{center}\end{minipage}\hfill
            \begin{minipage}[c]{0.47\textwidth}\begin{center}
              \colorbox{figColBox}{\includegraphics[trim=0mm 12mm 0mm 18mm,clip,width=.98\textwidth]{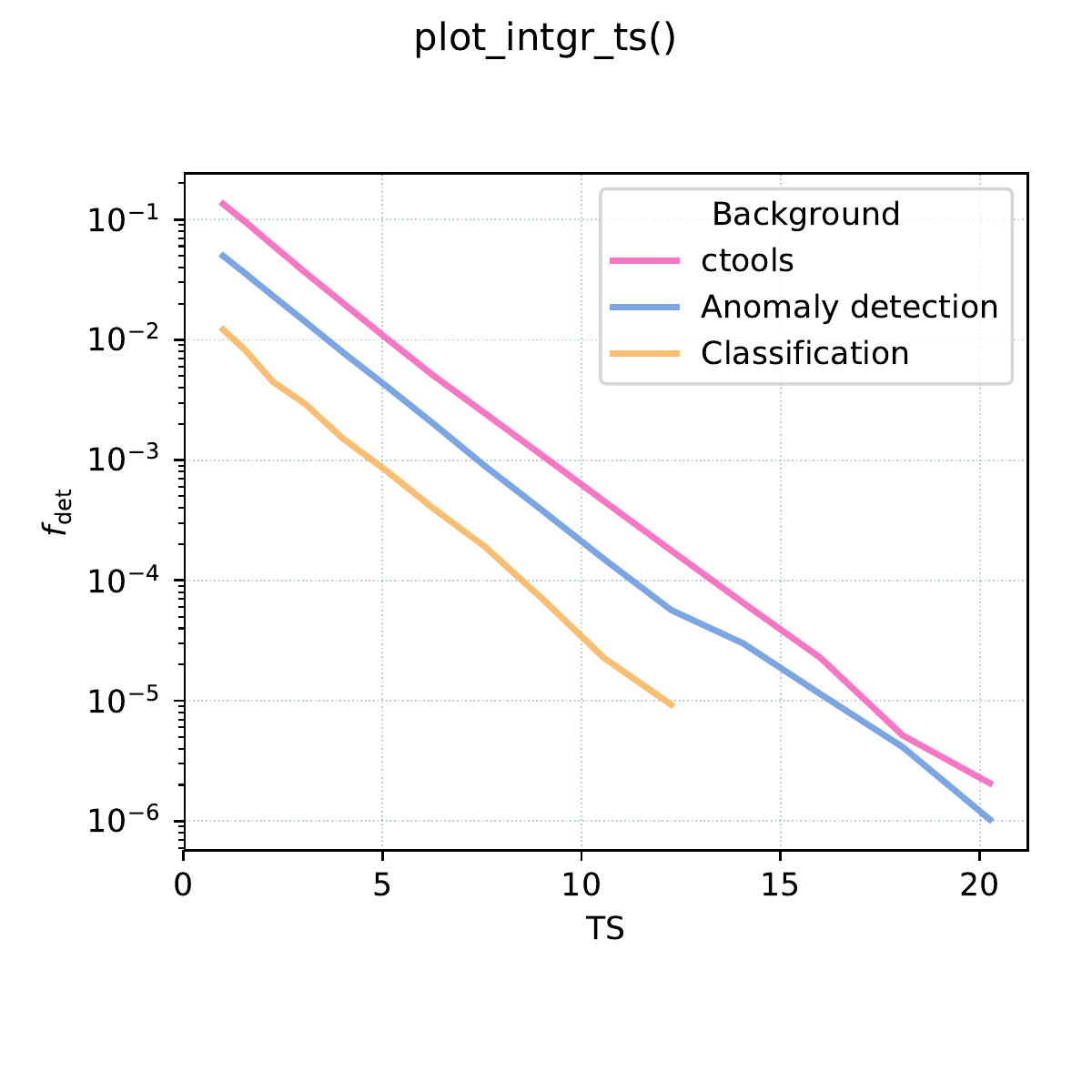}}
            \end{center}\end{minipage}\hfill 
          \end{center}
        \end{minipage}\hfill
        \vspace{5pt}
        \begin{minipage}[c]{1\textwidth}
          \begin{center}
            \begin{minipage}[c]{0.47\textwidth}\begin{center}
              \centering\subfloat[]{\label{FIGintgrTsSigSig}}
            \end{center}\end{minipage}\hfill
            \begin{minipage}[c]{0.47\textwidth}\begin{center}
              \centering\subfloat[]{\label{FIGintgrTsSigBck}}
            \end{center}\end{minipage}\hfill 
          \end{center}
        \end{minipage}\hfill
        %
        %
        \begin{minipage}[c]{1\textwidth}
          \begin{center}
            \begin{minipage}[t]{1\textwidth}\begin{center}
              \caption{\label{FIGintgrTsSig}Dependence
              of \fdet, the fraction of events with a \ts value
              larger than a given threshold,
              on the value of the threshold. The different detections methods
              are compared, derived for the
              signal~\Subref{FIGintgrTsSigSig}
              and background~\Subref{FIGintgrTsSigBck}
              samples, as indicated.
              The full lines in either figure correspond to the
              pre-trials test statistic.
              The dashed lines in~\Subref{FIGintgrTsSigSig}
              represent \fdet after accounting for trials,
              where in~\Subref{FIGintgrTsSigBck}
              we found ${\fdet(\ts>1) = 0}$ post-trials
              in all cases.
              For clarity, the relations are truncated to the
              range, ${1 < \ts < 100}$, where
              the pre-trials background distributions in~\Subref{FIGintgrTsSigBck}
              do not extend beyond ${\ts \approx 20}$.
              The dashed-dotted vertical
              line in~\Subref{FIGintgrTsSigSig} highlights
              the value, $\ts = 25$.}
            \end{center}\end{minipage}\hfill
          \end{center}
        \end{minipage}\hfill
        \vspace{15pt}
      \end{minipage}\hfill



      \begin{minipage}[c]{1\textwidth}

        \begin{minipage}[c]{1\textwidth}
          \begin{center}
            \begin{minipage}[c]{0.47\textwidth}\begin{center}
              \colorbox{figColBox}{\includegraphics[trim=0mm 12mm 0mm 18mm,clip,width=.98\textwidth]{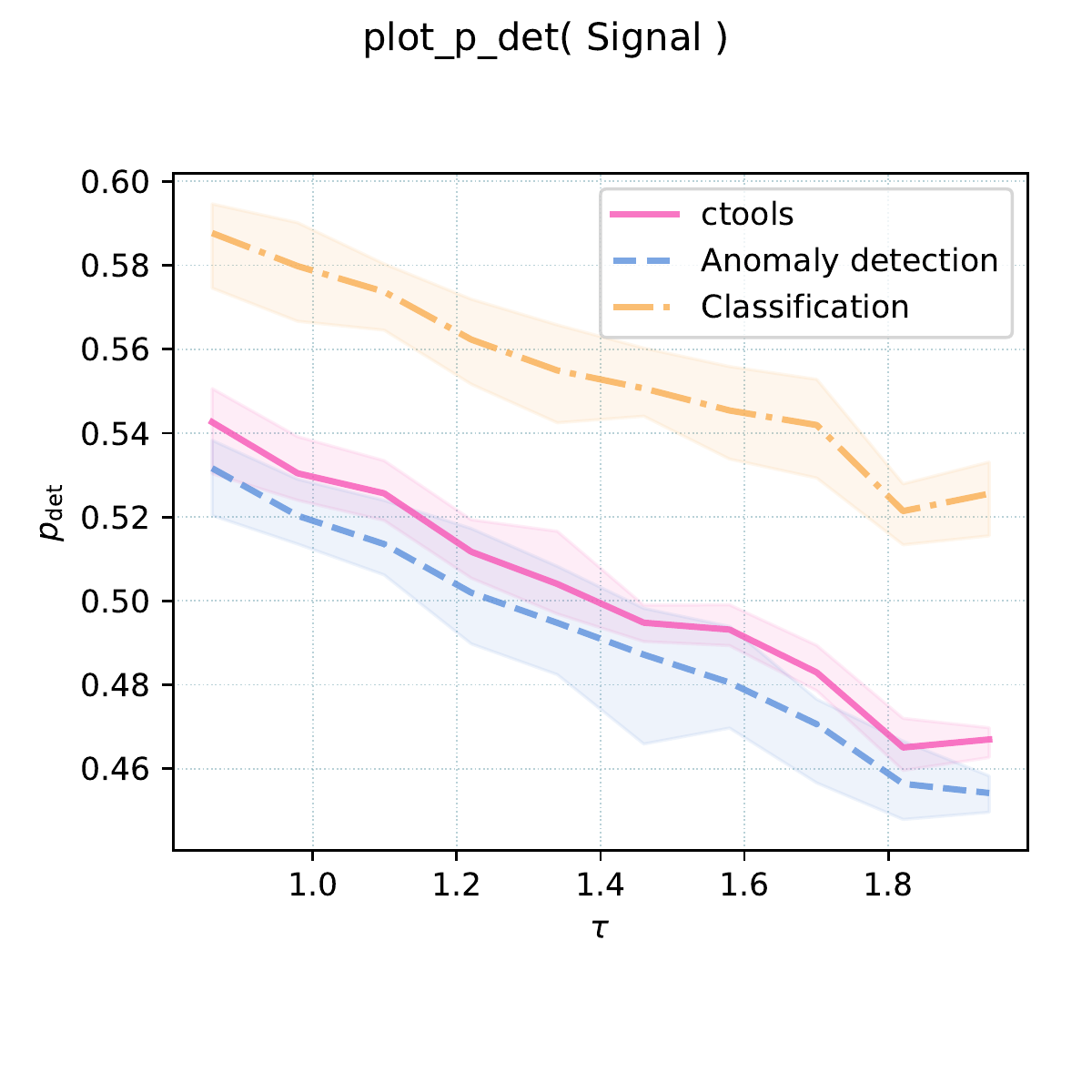}}
            \end{center}\end{minipage}\hfill
            \begin{minipage}[c]{0.47\textwidth}\begin{center}
              \colorbox{figColBox}{\includegraphics[trim=0mm 12mm 0mm 18mm,clip,width=.98\textwidth]{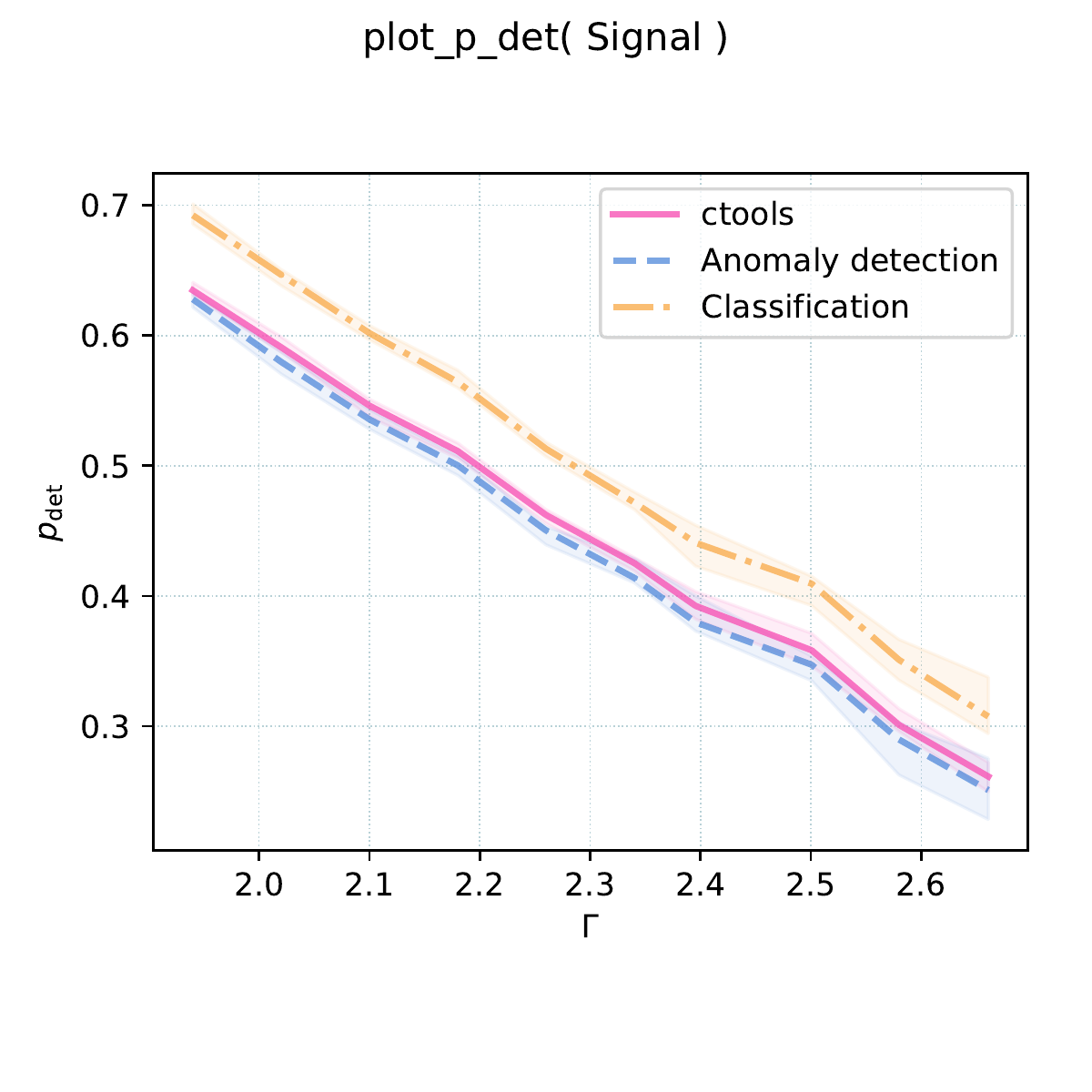}}
            \end{center}\end{minipage}\hfill 
          \end{center}
        \end{minipage}\hfill
        \vspace{5pt}
        \begin{minipage}[c]{1\textwidth}
          \begin{center}
            \begin{minipage}[c]{0.47\textwidth}\begin{center}
              \centering\subfloat[]{\label{FIGpdetTimeIndex}}
            \end{center}\end{minipage}\hfill
            \begin{minipage}[c]{0.47\textwidth}\begin{center}
              \centering\subfloat[]{\label{FIGpdetSpecIndex}}
            \end{center}\end{minipage}\hfill 
          \end{center}
        \end{minipage}\hfill
        %
        %
        \begin{minipage}[c]{1\textwidth}
          \begin{center}
            \begin{minipage}[t]{1\textwidth}\begin{center}
              \caption{\label{FIGpdetSpecIndexExCutoff}Dependence
              of \pdet on the temporal \Subref{FIGpdetTimeIndex}
              and spectral \Subref{FIGpdetSpecIndex} indices of simulated
              \llgrbs,
              after accounting for trials.
              The shaded regions correspond to~${1\sigma}$
              uncertainties on the values of \pdet, derived
              using the bootstrap method.
              The different detection methods are compared, as indicated.}
            \end{center}\end{minipage}\hfill
          \end{center}
        \end{minipage}\hfill
        \vspace{15pt}
      \end{minipage}\hfill

    \end{figure*}

    Figure~\ref{FIGintgrTsSig} shows 
    \fdet, the fraction of events with a \ts value
    larger than a given threshold,
    as a function of this threshold.
    We find that \ctools
    and the \andet achieve comparable
    significance distributions. The two methods
    enable detection of a similar
    fraction of the events,
    with slightly better performance by \ctools.
    The performance of the \cls approach
    is better than that
    of \ctools, with a relative improvement
    in detectability of~${\Sim10\%}$ on average.

    It is important to verify that the new detection
    methods do not produce spurious detections, and that
    the corresponding test statistics are properly mapped to
    significance. We therefore evaluate the different
    algorithms on the background sample,
    and compare them to the reference \ctools distribution.
    As shown in Figure~\ref{FIGintgrTsSigBck},
    the \andet and \cls methods produce
    comparable or better (lower) rates
    of fake detections.
    For the given sample of \powB{6} background simulations,
    none of the methods exceed a pre-trials \ts value of~$20$,
    or a post-trials value of~$1$.

    The dependence of \pdet
    on the temporal and spectral indices of the simulated \llgrbs
    is shown in Figure~\ref{FIGpdetSpecIndexExCutoff}.
    One may observe that our new methods
    match or improve upon the performance of \ctools.
    As expected, longer-lasting and harder spectra
    are more likely to be detected by all algorithms.

  \section{Summary}
    %
    In this study, we present a new approach for
    source detection.
    Our algorithm is based on deep learning, utilising
    recurrent neural networks, which are ideally suited
    for time series analyses.
    The model can be used to evaluate
    observation sequences
    of second time scales with insignificant latency.
    The choice of technology is therefore
    particularly fitting for real-time searches.

    We have developed two methods, 
    based on \andet and \cls techniques.
    \Andet represents a model-independent
    approach, where transient events are identified based on
    their divergence from the expected background.
    The method is data-driven. We thus
    avoid the need for
    background modelling, as well as for
    explicit characterisation of the state of the instrument.
    The \cls method allows one to perform
    targeted searches. In this case, the \rnn
    is trained to identify generic transient patterns.
    The estimator
    provides high detection rates while maintaining
    low fake rates.

  \acknowledgments
    %
    We would like to thank the following
    people for numerous useful discussions:
    D.\,Biehl,
    D.\,Boncioli,
    Z.\,Bosnjak,
    O.\,Gueta,
    T.\,Hassan,
    M.\,Krause,
    G.\,Maier,
    M.\,Nievas Rosillo,
    I.\,Oya,
    A.\,Palladino,
    E.\,Pueschel,
    R.\,R.\, Prado,
    L.\,Rauch,
    and
    W.\,Winter.

    This research made use of \ctools~\cite{Knodlseder:2016nnv}, a community-developed analysis package for Imaging Air Cherenkov Telescope data. \ctools is based on \gamlib, a community-developed toolbox for the high-level analysis of astronomical gamma-ray data.
    This research has also made use of the \cta \irfs (version \prodIrf), provided by the \cta Consortium and Observatory; see~\url{http://www.cta-observatory.org/science/cta-performance/} for more details.


\printbibliography

\end{document}